\documentclass[twocolumn]{pasj00}
\SetRunningHead{Astronomical Society of Japan}{A Study of Selection Methods for H$\alpha$ Emitting Galaxies at $z\sim1.3$}

\Received{2013/08/19}
\usepackage{times}
\usepackage{color}
\usepackage{url}

\begin{document}

\title{A Study of Selection Methods for H$\alpha$ Emitting Galaxies at $z\sim1.3$
for the Subaru/FMOS Galaxy Redshift Survey for Cosmology (FastSound)}
\author{Motonari \textsc{Tonegawa},\altaffilmark{1,2}
  Tomonori \textsc{Totani},\altaffilmark{1,2}
  Masayuki \textsc{Akiyama},\altaffilmark{3}
  Gavin \textsc{Dalton},\altaffilmark{4,11}
  Karl \textsc{Glazebrook},\altaffilmark{5}
  Fumihide \textsc{Iwamuro},\altaffilmark{2}
  Masanao \textsc{Sumiyoshi},\altaffilmark{2} 
  Naoyuki \textsc{Tamura},\altaffilmark{6}
  Kiyoto \textsc{Yabe},\altaffilmark{7}
  Jean \textsc{Coupon},\altaffilmark{8,9}
  Tomotsugu \textsc{Goto},\altaffilmark{10} and
  Lee R. \textsc{Spitler}\altaffilmark{5,12,13}
}
%\thanks{Last update: February 19, 2013}
\altaffiltext{1}{Department of Astronomy, The University of Tokyo,
  Bunkyo-ku, Tokyo 113-0033, Japan}
\email{tonegawa@astron.s.u-tokyo.ac.jp}
\altaffiltext{2}{Department of Astronomy, Kyoto University, 
  Sakyo-ku, Kyoto 606-8502, Japan}
\altaffiltext{3}{Astronomical Institute, Tohoku University, Aoba-ku, Sendai, Miyagi 980-8578}
\altaffiltext{4}{Department of Physics, University of Oxford, Keble Road, Oxford, OX1 3RH, UK}
\altaffiltext{5}{Centre for Astrophysics \& Supercomputing, Swinburne University of Technology, P.O. Box 218, Hawthorn, VIC 3122, Australia}
\altaffiltext{6}{Kavli Institute for the Physics and Mathematics of the Universe, The University of Tokyo}
\altaffiltext{7}{National Astronomical Observatory of Japan, 2-21-1 Osawa, Mitaka, Tokyo 181-8588}
\altaffiltext{8}{Astronomical Observatory of the University of Geneva, ch.\ d'Ecogia 16, 1290 Versoix, Switzerland}
\altaffiltext{9}{Institute of Astronomy and Astrophysics, Academia Sinica, P.O. Box 23-141, Taipei 10617, Taiwan}
\altaffiltext{10}{Dark Cosmology Centre
Niels Bohr Institute, University of Copenhagen
Juliane Maries Vej 30, 2100 Copenhagen O, Denmark}
\altaffiltext{11}{STFC RALSpace, Harwell Oxford, OX11 0QX, UK}
\altaffiltext{12}{Australian Astronomical Observatory, P.O. Box 296 Epping, NSW 1710, Australia}
\altaffiltext{13}{Department of Physics \& Astronomy, Macquarie University, Sydney, NSW 2109, Australia}

\maketitle

\begin{abstract}
  The efficient selection of high-redshift emission galaxies is
  important for future large galaxy redshift surveys for cosmology.
  Here we describe the target selection methods for the FastSound
  project, a redshift survey for H$\alpha$ emitting galaxies at
  $z=1.2$--$1.5$ using Subaru/FMOS to measure the linear growth rate
  $f\sigma_8$ via Redshift Space Distortion (RSD) and constrain the
  theory of gravity. To select $\sim\!400$ target galaxies in the
  $0.2$ deg$^2$ FMOS field-of-view from photometric data of
  CFHTLS-Wide ($u^{*}g'r'i'z'$), we test several different methods
  based on color-color diagrams or photometric redshift estimates from
  spectral energy distribution (SED) fitting.  We also test the
  improvement in selection efficiency that can be achieved by adding
  near-infrared data from the UKIDSS DXS ($J$).  The success rates of
  H$\alpha$ detection with FMOS averaged over two observed fields using these methods are
  $11.3\%$ (color-color, optical), $13.6\%$ (color-color,
  optical+NIR), $17.3\%$ (photo-$z$, optical), and $15.1\%$
  (photo-$z$, optical+NIR). Selection from photometric redshifts tends
  to give a better efficiency than color-based methods, although
  there is no significant improvement by adding $J$ band data within
  the statistical scatter.  We also investigate the main limiting
  factors for the success rate, by using the sample of the HiZELS
  H$\alpha$ emitters that were selected by narrow-band imaging.
  Although the number density of total H$\alpha$ emitters
  having higher H$\alpha$ fluxes than the FMOS sensitivity is
  comparable with the FMOS fiber density, the limited accuracy of
  photometric redshift and H$\alpha$ flux estimations have comparable effects
  on the success rate of $\lesssim 20\%$ obtained from SED fitting.
\end{abstract}

\section{Introduction}
Recent cosmological observations, including distance measurements of
Type Ia supernovae, temperature fluctuations of the Cosmic Microwave
Background (CMB), and large galaxy redshift surveys, have revealed that
the expansion of the Universe is accelerating. This phenonemon cannot be
explained by conventional physics (see e.g., \cite{Peebles};
\cite{Frieman}; \cite{Weinberg} for recent reviews).  This
acceleration may be caused by the existence of an unknown form of
energy (so-called ``dark energy'') coming into the energy-momentum
tensor in the Einstein's field equation. Another possibility is that
the general relativity is not the correct theory of gravity on
cosmological scales.
 
Observations of the large scale structure of the universe as traced by
large spectroscopic galaxy survey are now attracting significant
attention as a probe to investigate the nature of dark energy.  For
example, the baryon acoustic oscillation (BAO) signature found in the
clustering pattern can be used as a standard ruler to measure the
geometry of the universe and then to constrain the equation of state
of the dark energy (\cite{Cole}; \cite{Eisenstein2005};
\cite{BlakeBAO}; \cite{BeutlerBAO}; \cite{Anderson}).  Galaxy redshift
surveys also provide a useful test of the gravity theory on
cosmological scales (\cite{Hawkins}; \cite{Guzzo}; \cite{Blake};
\cite{Samushia}; \cite{Reid}; \cite{BeutlerRSD}; \cite{delaTorre}),
because the amplitude of the redshift space distortion (RSD) induced by peculiar
motions is related to the structure growth rate (\cite{Kaiser};
\cite{Hamilton}).

The FastSound
project\footnote{\url{http://www.kusastro.kyoto-u.ac.jp/Fastsound/}} is a
redshift survey of H$\alpha$ emitting galaxies at $z \sim 1.3$, aiming
at the first significant detection of RSD beyond $z = 1$. The survey
utilizes the Fiber Multi-Object Spectrograph (FMOS, \cite{Kimura}) on
the 8.3m Subaru Telescope, which can observe 400 simultaneous
near-infrared spectra from within a 30 arcmin diameter field-of-view
(FoV).  The survey started in March 2012, and will collect in total $\gtrsim
5,000$ galaxy redshifts in four fields with a total survey area of
$30\;{\rm deg^2}$ in two years. The main science goal is to measure
the growth rate $f\sigma_8$, where $f={\rm dln}D/{\rm dln}a$ is the
evolution of the growth factor $D$ with cosmic scale factor $a$ and
$\sigma_8$ is the total amplitude of matter fluctuations. The
estimated accuracy of the RSD measurement is $\sim 15\%$,\footnote{\url{http://www.
subarutelescope.org/Science/SACM/Senryaku/FMOS_Cosmology_proposal.pdf}} which would
give a stronger constraint on the theory of gravity than currently
available from other RSD surveys at $z<1$.

A key requirement for the FastSound survey design is to determine how
to select the target galaxies efficiently from photometric data.  To
finalize the target selection method for FastSound, we carried out
test observations using FMOS in September/October, 2011, where we
tested several selection methods, e.g., color-based selections versus
photometric-redshift-based selections, and pure optical-band
selections versus adding near-infrared bands. The aim of this paper is
to report the results of this pilot study.  Though the
objective here is mainly for the FastSound project, the results will also
be useful for efficient selection of emission-line galaxies or
star-forming galaxies at high redshifts, which can aid the 
design of even larger future cosmological surveys.  We will also
examine the factors limiting the selection efficiency by comparing our
results with the statistics of emission line galaxies detected by
narrow-band imaging from the HiZELS survey (\cite{Geach}; \cite{Sobral}).

This paper is organized as follows.  In \S\ref{sec:target_selection},
we describe the photometric data used to select targets, the methods of
estimating photometric redshifts and H$\alpha$ fluxes, and the selection
methods tested. In \S\ref{sec:observation}, the data reduction is
presented briefly, followed by the description of the automated
emission line detection method used. The main results are
presented in \S\ref{sec:results}, and we summarize our results in
\S\ref{conclusion}.
Throughout this paper, all magnitudes are given in the AB system, and 
a standard cosmology of $(\Omega_m,\Omega_\Lambda,h)=(0.3,0.7,0.7)$ is adopted.

\section{Target Selection for FMOS Observation}
\label{sec:target_selection}
\subsection{Input Photometric Galaxy Catalogues for Selection}
\label{sample}
For the FastSound project, we must efficiently select target galaxies
which are detectable by FMOS in an exposure time of 30 min (H$\alpha$
emission line fluxes of $F_{\rm H\alpha}\gtrsim1.0\times10^{-16}\;{\rm
  [erg/cm^2/s]}$ at $z=1.2$--$1.5$) in a survey area of $\sim 30\;
{\rm deg^2}$. We adopted the Canada-France-Hawaii Telescope Legacy
Survey (CFHTLS) Wide as the primary photometric data set for the
target selection. We used the MAG\_AUTO magnitudes of $u^{*}, g', r',
i'$, and $z'$ filters of the $z'$-selected merged catalogue in the
CFHTLS-Wide W1 and W4 fields generated by \citet{Gwyn} for these test
observations.  The approximate limiting magnitudes are $26.0$, $26.5$,
$25.9$, $25.7$, and $24.6$ respectively ($50\%$ completeness level for
point sources).

In $\sim 3 \;{\rm deg^2}$ of the CFHTLS W1 field and $\sim 7 \;{\rm
  deg^2}$ of the W4 field, near-infrared $J$ band data are also
available from the UKIDSS deep Extragalactic Survey (UKIDSS DXS:
\cite{Lawrence}) DR8, whose limiting magnitude is $22.3$ ($5\sigma$,
point-source), and we use these data to test the usefulness of adding
NIR data to the selection criteria. The Galactic extinction was
corrected using the $E(B-V)$ maps of \citet{Schlegel}.

\subsection{Color-based Selection}
\label{color}

\begin{figure*}
   \begin{center}
      \FigureFile(80mm,80mm){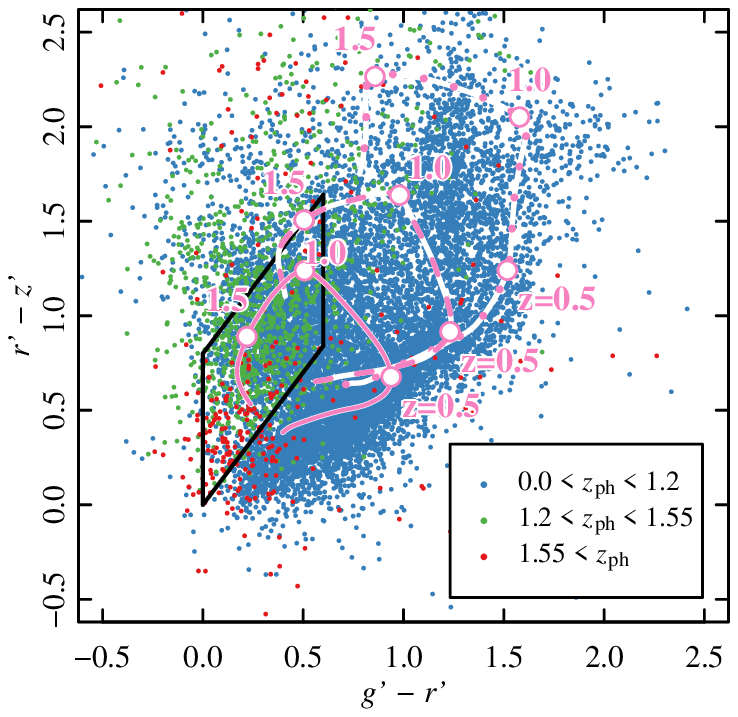}
      \FigureFile(80mm,80mm){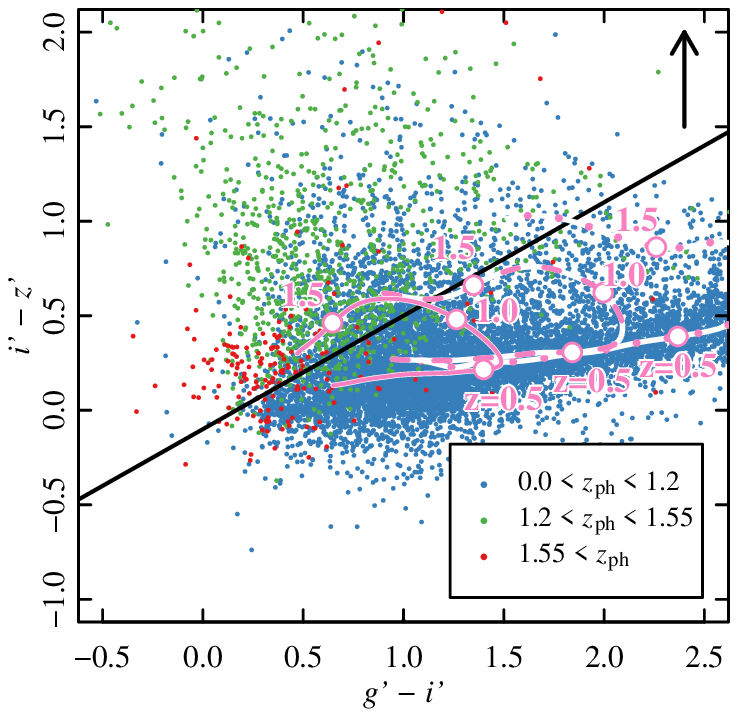}
   \end{center}
   \caption{CFHTLS wide galaxies on the $g'r'z'$ and $g'i'z$ diagrams. 
The plots are color-coded according to the photometric redshifts,
as indicated in the legend.
The black lines indicate the color selection criteria of
$0.0<(r'-z')-1.4(g'-r')<0.8$, $0.0<g'-r'<0.6$, 
and $(i'-z')-0.6(g'-i')>-0.1$.
Magenta lines show the color tracks of unreddened elliptical galaxies (dotted), Sbc galaxies (dashed),
and Scd galaxies (solid) at $0.0<z<2.0$.
These expected colors are based on SEDs from Coleman et al. (1980).}
\label{twocolor}
\end{figure*}

\begin{figure}
   \begin{center}
      \FigureFile(80mm,80mm){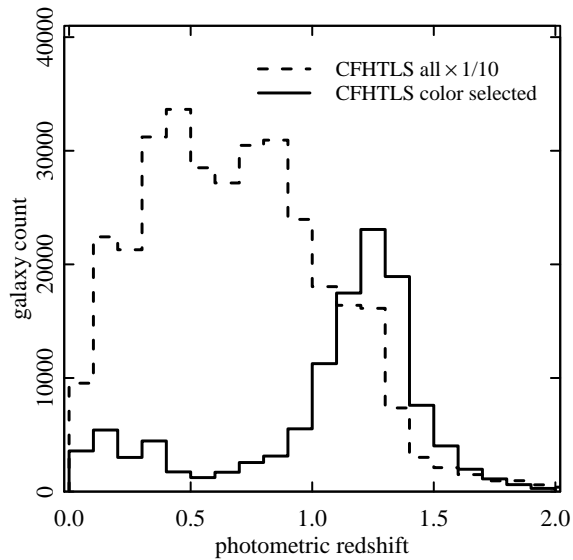}
   \end{center}
\caption{Photo-$z$ distributions of all galaxies (\textit{solid})
and color-selected galaxies (\textit{dashed}) of CFHTLS W1 and W4.
The wavelength coverage of FastSound corresponds to $z=[1.19,1.53]$.
}
\label{photoz_distrib}
\end{figure}

The first simple method that we test is a selection by galaxy
colors. We empirically determined the color selection criteria to
select galaxies whose H$\alpha$ fluxes are expected to be strong,
based on H$\alpha$ flux estimates from SED fittings and photometric
redshift calculations using deep optical+NIR photometry of galaxies in
the Subaru deep Field and the Subaru XMM-Newton deep Field (Sumiyoshi
et al. 2009).  For this observational study we test the following
color selection criteria. In the case of selection using only optical
photometry of CFHTLS wide, we adopt: $0.0<(r'-z')-1.4(g'-r')<0.8$,
$0<g'-r'<0.6$, $(i'-z')-0.6(g'-i')>-0.1$. These conditions are shown
in color-color diagrams of Figure \ref{twocolor}.  In addition to
these color conditions, we also adopt a magnitude range of
$22.0<g'<23.8$, which is empirically determined from past FMOS
observations.$^2$ In this figure the galaxies are color-coded
according to photometric redshifts described in \S\ref{estimation}.
Approximately one third of galaxies with $z_{ph} = 1.2$--$1.5$ that we
want to observe are populated in the regions of the above conditions,
while more than 90\% of the galaxies outside this redshift range are
rejected.  We varied the magnitude thresholds of $z'$ in a range of $z
\sim 22.6$--$23.0$ so that $\sim\!400$ target galaxies are available
for a FoV (see Table \ref{rate} for values in each field).  

The photo-$z$ distribution of color-selected
galaxies is displayed in Figure \ref{photoz_distrib}.  The
distribution of all galaxies found in the primary catalog is
overlaid.  The distribution of selected galaxies indeed has its peak
at around $z\sim1.3$, though there still remain low-redshift
galaxies that would be a contamination in our target sample.

For the test of color selection using the NIR $J$ band, we further
added a condition of $i'-J > 0.5$, to account for the redshifted
$4000$ \AA \ break falling between $i'$ and $J$.  However, we found
that this condition was too strong to retain $\sim\! 400$ targets in
the FMOS FoV for the W4 field, and hence the condition was relaxed to
$i'-J > 0.0$ (see \S \ref{sec:observation}).

\subsection{Selection by Photometric Redshifts and H$\alpha$ Flux Estimation}
\label{estimation}
The second method we tested is based on photometric redshift estimates
and H$\alpha$ flux estimates from SED fitting. For galaxies brighter
than $i'=24$, the CFHTLS T0006 official photometric redshifts are
available (\cite{Ilbertphotoz}, \cite{Coupon}), and we use these. The
redshift accuracy $\sigma_{\Delta z/(1+z_s)}\equiv 1.48\times
\mathrm{median}(\left|\Delta z\right|/(1+z_s))$ is around $0.035$ at
$i'<21.5$, $0.042$ at $21.5<i'<22.5$, and $0.070$ at $22.5<i'<23.5$.
However, no photometric redshift estimate is provided for galaxies
fainter than $i'=24$ due to the poor reliability estimates of the
photometry.  Here, since we wish to increase the chance of observing
$z > 1$ galaxies, we computed photometric redshifts for galaxies at
$i'>24$ by ourselves using the public code {\it LePhare}
(\cite{Arnouts}, \cite{Ilbertphotoz}) according to the description of
Coupon et al. (2009).  When a galaxy is not detected in one of the
band filters, the band was simply not used in the photo-$z$
calculation.

Photometric zero-point calibration using galaxies with known
spectroscopic redshifts is important for the accuracy of photometric
redshift calculation. Since the spectroscopic data of the VVDS
(VIMOS-VLT Deep Survey: \cite{Febre};
\cite{Garilli})\footnote{\url{http://cesam.oamp.fr/vvdspub/}} deep and
wide survey data are available on the area of CFHTLS W1 and W4
respectively, we used them to perform the zero-point calibration of
CFHTLS wide photometry, although they are outside the redshift range
of our survey.  The VVDS survey depths are $17.5\leq I_{AB}\leq 24.0$
for VVDS deep and $17.5\leq I_{AB} \leq 22.5$ for VVDS wide.  We used
galaxies with flags of 2--4 ($\sim$ 80--99\% secure redshifts),
  $9$ (one secure spectral feature in emission), $12$--$14$, or $19$
  (similar to 2--4, 9 but flags for broad line AGNs) for both VVDS
deep and wide to assure secure redshift determination.
Galaxies from VVDS deep and wide public data with $z'$ magnitudes brighter than
    $23.0$ are matched with CFHTLS W1 and CFHTLS W4, respectively, and
    used for the derivation of photometric calibrations.  The sample
    sizes of VVDS-CFHTLS are $5{,}054$ (CFHTLS W1) and $8{,}445$ (W4),
    and the redshift medians are $0.69$ and $0.56$, respectively.

These photometric redshifts are based on empirical SED templates of
Ell, Sbc, Scd, Irr (CWW; \cite{Coleman}), $60$ interpolated SEDs from
CWW, and two starbursts from Kinney et al.~(1996). H$\alpha$ fluxes
cannot be estimated simply from these results, because the empirical
templates do not include any details on the physical galaxy properties
such as stellar mass or star formation rate.  Therefore we further
performed fitting with theoretical SED templates from PEGASE2
(\cite{Fioc}), Scalo IMF (\cite{Scalo}) and solar abundance, fixing
the redshifts at those estimated using the empirical SED templates. We
use $13$ models of exponentially declining star formation histories
with the exponential time scales ranging $0.1$--$20$ Gyr, and a
constant SFR model. The dust extinction is taken into account assuming
the Calzetti law (Calzetti et al. 2000) in the range of $E(B-V) =
0$--$0.35$. Then we calculate the intrinsic H$\alpha$ flux from SFR by
the conversion factor which is based on Kennicutt (1998) and of which the offset
is calibrated using SDSS galaxies (\cite{Brinchmann}, \cite{Sumiyoshi}).
Finally, the observable
H$\alpha$ flux is calculated taking into account extinction using the
$E(B-V)$ value obtained from the SED fit. Here, we followed the
prescription of Cid Fernandes et al.~(2005) to take into account that
the extinction of H$\alpha$ flux (from hot ionized gas region) is
generally higher than that of stellar radiation estimated from SED
fittings.

For the photo-$z$ selection using optical bands, first we selected
target galaxies at $1.18<z_{ph}<1.54$, $z'<23.0$ and $g'-r' < 0.7$.
The constraints on $z'$ magnitude and color are added empirically,
since we learned from the past FMOS observations that bluer galaxies
have higher probability of detectable emission lines, and that
galaxies fainter than $z' = 23$ are not usually detected at our survey
limits. The dependence of the selection efficiency on the
  $g'-r'$ threshold will be discussed in \S\ref{detectionrate}.

Then we select galaxies with bright estimated H$\alpha$
fluxes, with a flux threshold determined to keep $\sim\! 400$ target
galaxies for one FMOS field. The typical threshold flux is $\sim\!0.75
\times 10^{-16}\;{\rm [erg/cm^2/s]}$
(see Table \ref{rate} for the values in each field).

In the case of photo-$z$ selection using NIR $J$ band, we repeated the
same procedures as above, except that we calculated photo-$z$'s for
all the galaxies because the CFHTLS official photometric redshifts are
based only on the optical bands. It should be noted that only
  about 70\% of all the CFHTLS Wide galaxies are detected in the
  UKIDSS $J$ band, and the fraction in the galaxies selected as the
  FMOS target is also similar.  Photo-$z$'s of the galaxies not
  detected in $J$ are calculated only with the optical bands, and this
  may limit the improvement of the line detection rate by adding a NIR
  band.  We will also test the effect of including the $K$ band on the
  target selection efficiency in \S\ref{detectionrate}.

Finally, it should also be noted that our target
selection does not include particular conditions to remove
AGNs. Therefore we expect a contamination of AGNs in our sample, but
examination of line widths of detected galaxies by our FMOS
observation indicates that the fraction is not large.
  
\begin{table*}
\begin{center}
  \caption{The positions of the FMOS field centers, selection methods, and
    Galactic extinction values for our observations. }
\begin{tabular}{cccccc}
\hline
\hline
Field  & RA(J2000) & DEC(J2000) & selection & band & mean E(B-V)\\
\hline
W1\_001 & 02:25:51.85 & $-$04:27:00.0 & color & optical & 0.026\\
W1\_002 & 02:27:35.77 & $-$04:27:00.0 & color & optical+NIR & 0.027\\
W1\_003 & 02:05:10.00 & $-$04:49:30.0 & photo-$z$ & optical & 0.027\\
W1\_004 & 02:26:44.00 & $-$04:49:30.0 & photo-$z$ & optical+NIR & 0.027\\
\hline
W4\_001 & 22:14:03.96 & $+$01:40:30.0 & color & optical & 0.038 \\
W4\_002 & 22:14:55.92 & $+$01:18:00.0 & color & optical+NIR & 0.041 \\
W4\_003 & 22:15:54.00 & $+$01:40:30.0 & photo-$z$ & optical & 0.044 \\
W4\_004 & 22:16:40.00 & $+$01:18:00.0 & photo-$z$ & optical+NIR & 0.046 \\
\hline
\hline
\end{tabular}
\label{fovs}
\end{center}
\end{table*}

\section{FMOS Observation}
\label{sec:observation}

We observed four FMOS fields for each of CFHTLS W1
(hereafter we call them W1\_001--004) and W4 (W4\_001--004).  The
four different combinations of the used photometric bands (optical
only or optical+NIR) and selection methods (color or photometric
redshift) are applied to the four fields in each of the two CFHTLS
wide fields. The field locations and corresponding selection parameters
are summarized in Table \ref{fovs}.

\subsection{Observation and Data Reduction}\label{reduction}
The observations were carried out with Subaru/FMOS on the nights of
September 22--25$^{\rm th}$, 2011.  We used the normal beam-switching
mode (NBS mode) of FMOS, with the total exposure time of $90$ min in
one FoV for each of the object and sky frames. In the NBS
mode, the object frame is taken using all 400 fibers for targets,
followed by the sky frame. We chose this mode because the fiber
allocation is geometrically less complicated than the cross beam
switching (CBS), where 400 fibers are split into 200 targets and 200
sky regions, and sky subtraction is done by another frame exchanging
target and sky fibers.\footnote{\url{http://www.naoj.org/Observing/Instruments/FMOS/}}

FMOS has two different spectral resolution modes, and
the FastSound project uses the high-resolution (HR) mode ($R\sim2200$)
since the throughput in HR mode is higher than in the
low-resolultion (LR) mode by a factor of about two due to loses in the
low resolution (LR) mode arising from the additional Volume-Phase
Holographic grating used to decrease spectral resolution
(\cite{Tamura}).  However, in this observing run, the HR mode was not
available for IRS1 due to a mechanism failure, and hence the
IRS1 observation was done with the LR mode covering
$0.9\;\mu$m--$1.8\;\mu$m ($R\sim500$). 
For this reason, we only present the
results from IRS2 in this paper, i.e., $\sim\!200$ galaxies in one FoV.
The exposure time was set to be three times longer than that for 
FastSound project, because the background of IRS2 was $\sim2$ times higher
than usual.

For IRS2, the wavelength coverage of the HR mode was set to be at the
blue-end of $H$ band, $1.44 \;\mu$m--$1.66 \;\mu$m (corresponding to
H$\alpha$ at $z=1.19$--$1.53$), and the pixel scale was $1.07\;{\rm
  [\AA/pix]}$. This range was chosen to be optimal, considering the
balance between the maximal scientific value of FastSound (i.e.,
higher redshifts) and a sufficient number of emission line galaxies
detectable by FMOS, based on the results of previous preliminary 
observations.  The obtained data were reduced with the FMOS data
reduction pipeline (\textit{FIBRE-pac}: Iwamuro et al. 2012).

\subsection{Emission Line Detection}\label{detection}
Emission lines were searched for by automated emission line
detection software developed by the FastSound group (Tonegawa et al.
2013, in preparation), and here we briefly summarize the software
algorithm.  First we subtract the continuum component from each
spectrum in 2D images by polynomial fits. Then the 2D images of the spectra are
convolved with 2D Gaussian kernel, with standard deviations of
$\sigma_x = 4.26\;{\rm [pix]}$ in the wavelength direction
(corresponding to a quadratic sum of the FMOS HR mode resolution and a
typical velocity dispersion 175 km/s FWHM of H$\alpha$ lines detected
by FMOS at $z \sim 1.3$) and $\sigma_y = 2.5\;{\rm [pix]}$ for the
direction perpendicular to wavelength (corresponding to the
instrumental image quality).  The line flux is calculated by fitting
the kernel shape, and corresponding statistical noise is also
calculated from the noise map produced by {\it FIBRE-pac}.  A key
feature is that we artificially amplify the noise level in the regions
corresponding to the OH-suppression masks to effectively reduce the
signal-to-noise ($S/N$) ratio of the OH-airglow emission regions, where
the large scatter of $S/N$ causes false detections.

\begin{figure*}[t]
 \begin{center}
 \FigureFile(80mm,80mm){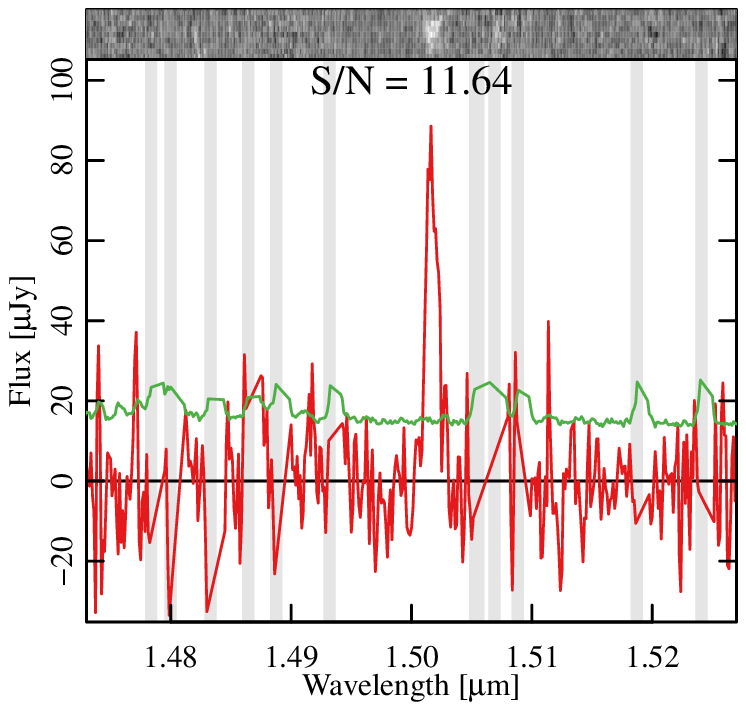}
 \FigureFile(80mm,80mm){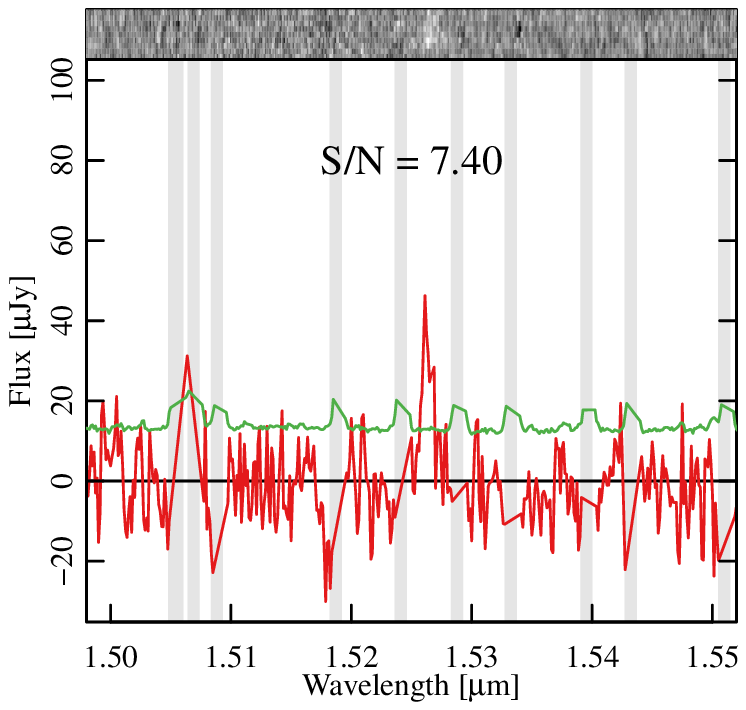}
 \FigureFile(80mm,80mm){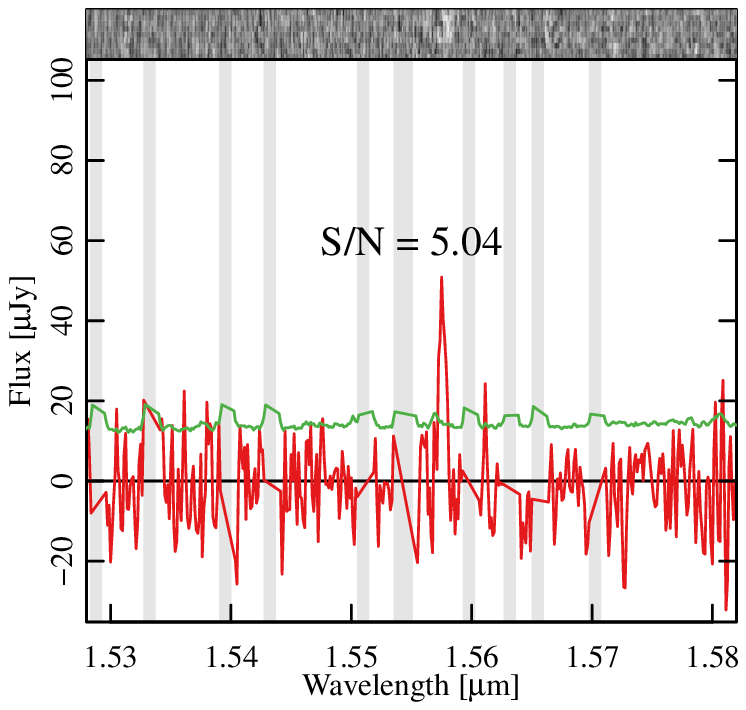}
 \FigureFile(80mm,80mm){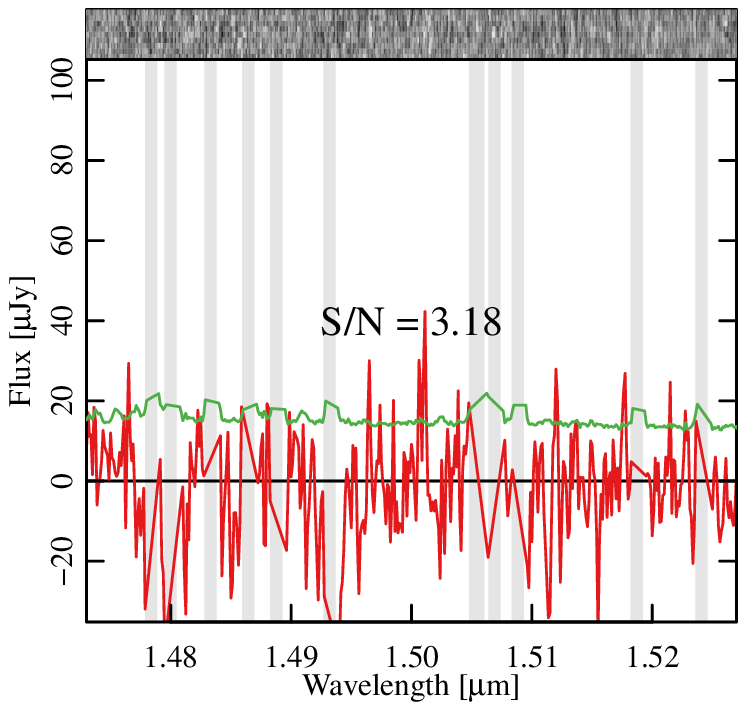}
 \caption{ Examples of 2D images (at the top-end of each panel) and
   1D spectra (main figures of each panel) of four emission line
   candidates with different $S/N$ values. Gray vertical stripes indicate the
   positions of OH-airglow suppression mask. Green lines show the
   noise level, multiplied by a factor of 5 for presentation
   purposes. These objects were examined by eye and judged to be the correct identifications for these
   objects, with the exception of the lowest S/N object in the lower right panel, which was rejected.
 }
 \label{1d2d}
 \end{center}
\end{figure*}

The algorithm then searches for local $S/N$ peaks along the wavelength
direction, and selects the local peaks having $S/N$ values larger than
a given threshold as the emission line candidates.  Finally, spurious
detections are removed by using information from the bad pixel and
noise maps. We examined the performance of the line detection software
by applying it to the inverted image produced by exchanging the object
and sky frames, where all the detections must be spurious. We find
that practically all emission lines are real above the threshold value
of $S/N > 4.5$ that was adopted in the sample presented in this work.
(Details will be reported in Tonegawa et al. 2013, in preparation.)
Examples of emission line candidates detected by the software are
shown in Figure \ref{1d2d}.

Since H$\alpha$ is the strongest line in star-forming galaxies in the
interested wavelength, the majority ($\gtrsim$ 90\%) of the detected emission
lines are expected to be H$\alpha$ (\cite{Glazebrook}, \cite{Yabe}).
In the following
we present the success rate of emission line detection, 
without discriminating between H$\alpha$ and other lines. 
We will discuss the contamination fraction of other lines
later in \S\ref{hizels} using the HiZELS sample.

We measure redshifts and H$\alpha$ fluxes by fitting a Gaussian
profile.  The effect of fiber losses are estimated from the relation
between the covering fraction (of light falling within the $1.2''$
diameter fiber aperture) and half-light radius of galaxies used in
Yabe et al. (2012).
This estimate was made for FMOS fields in which the NIR
data was available (i.e., W1\_002, W1\_004, W4\_002, W4\_004),
and the half-light radius was estimated from the intrinsic
radius of UKIDSS DXS galaxies in $J$-band filter and the typical seeing size of our
observation ($0.65''$ for W1 and $0.8''$ for W4, FWHM).  The intrinsic
radius was estimated by subtracting WFCAM seeing ($0.8''$ FWHM) from
the observed size (mean of the major and minor axis sizes) in
quadrature.  The median of covering fraction are $0.48$ for W1 and
$0.46$ for W4, respectively, and all line fluxes of detected objects
are uniformly corrected by these values.

\section{Results and Discussion}
\label{sec:results}

\begin{table*}[t]
\footnotesize
\begin{center}
\caption{Success rate statistics in each FoV. See
  \S\ref{sec:target_selection} for the detailed criteria.  (Note that
  the $i'-J$ condition is different for W1\_002 and W4\_002.)  In
  addition to our baseline statistics with variable $z'$ or $F_{\rm
    H\alpha, est}$ thresholds to keep $\sim$400 targets in one field,
  we also show the supplementary statistics when the uniform
  thresholds are adopted to all the fields.  The error bars on success
  rates are based on Poission statistics. }
\begin{tabular}{cccccccc}
\hline
\hline
FoV & Method & Band  & Target & Emitters & SR ($\%$) & $z'$ threshold & $F_{\rm H\alpha, est}$ threshold [erg/cm$^2$/s] \\
\hline
\multicolumn{8}{c}{all observed objects} \\
\hline
W1\_001 & color & opt & $179$ & $20$ & $11.2\pm2.5$ & $22.65$ & -- \\
W1\_002 & color & opt+NIR & $178$ & $32$ & $18.0\pm3.2$ & $22.90$ & -- \\
W1\_003 & photo-z & opt & $171$ & $29$ & $17.0\pm3.1$ & $22.90$ & -- \\
W1\_004 & photo-z & opt+NIR & $172$ & $27$ & $15.7\pm3.0$ & $23.05$ & -- \\
\hline
W4\_001 & color & opt & $174$ & $20$ & $11.5\pm2.6$ & -- & $0.85\times 10^{-16}$ \\
W4\_002 & color & opt+NIR & $174$ & $16$ & $9.2\pm2.3$ & -- & $0.80\times 10^{-16}$ \\
W4\_003 & photo-z & opt & $181$ & $32$ & $17.7\pm3.1$ & -- & $0.75\times 10^{-16}$ \\
W4\_004 & photo-z & opt+NIR & $164$ & $24$ & $14.6\pm3.0$ & -- & $0.65\times 10^{-16}$ \\
\hline
\multicolumn{8}{c}{uniform threshold of $z'<22.65$ (for color-selection)} \\
\hline
W1\_001 & color & opt & $179$ & $20$ & $11.2\pm2.5$ & $22.65$ & -- \\
W1\_002 & color & opt+NIR & $123$ & $22$ & $17.9\pm2.6$ & $22.65$ & -- \\
W4\_001 & color & opt & $171$ & $29$ & $17.0\pm3.1$ & $22.65$ & -- \\
W4\_002 & color & opt+NIR & $161$ & $25$ & $15.5\pm2.9$ & $22.65$ & -- \\
\hline
\multicolumn{8}{c}{uniform threshold of $F_{\rm H\alpha, est}>0.85\times 10^{-16}\;{\rm [erg/cm^2/s]}$ (for photoz-selection)} \\
\hline
W1\_003 & photo-z & opt & $116$ & $15$ & $12.9\pm2.2$ & -- & $0.85\times 10^{-16}$ \\
W1\_004 & photo-z & opt+NIR & $90$ & $9$ & $10.0\pm1.7$ & -- & $0.85\times 10^{-16}$ \\
W4\_003 & photo-z & opt & $173$ & $31$ & $17.9\pm3.1$ & -- & $0.85\times 10^{-16}$ \\
W4\_004 & photo-z & opt+NIR & $134$ & $20$ & $14.9\pm2.7$ & -- & $0.85\times 10^{-16}$ \\
\hline
\hline
\end{tabular}
\label{rate}
\end{center}
\end{table*}

\subsection{Statistics of Line Detection}\label{detectionrate}

The statistic that we are primarily interested in is the emission line
detection rate among the galaxies observed by FMOS (hereafter the
"success rate" or "SR"), for various selection methods.  This statistics is
summarized in Table \ref{rate} for each of eight FMOS FoVs observed.
The typical success rate is $\sim\!10$--$20\%$, and a general tendency
of higher detection rate by the photo-$z$ selections, relative to the
color selections, can be seen. On the other hand, we can see no clear
increase of the success rate, within the statistical uncertainties, due
to adding NIR data.  An increase of the success rate with added NIR
data can be seen for the case of color selection in the W1 field, but
it cannot be seen in the W4 field or in the cases of photo-$z$
selections.  It was expected that NIR data would be useful because of
the inclusion of redshifted 4000\AA \ breaks in photo-$z$
calculations, but this effect may not be large for blue, star-forming
galaxies having rather weak breaks. 

The thresholds of $z' < 22.6$
(color selection) or $F_{\rm H\alpha, est}$ (photo-$z$ selection) are
different for different fields, to keep $\sim$400 objects in each FMOS
FoV.  In order to examine whether the lower success rate (such as
W4\_002) is a result of a fainter $z'$ mag or $F_{\rm H\alpha, est}$
threshold in a field, we limit the statistics by taking the tightest
ones of $z'<22.65$ or $F_{\rm H\alpha, est}>0.85\times10^{-16}\;{\rm
  [erg/cm^2/s]}$ uniformly for all fields, which is also summerized in
Table \ref{rate}.  The success rates are almost unchanged, showing
that the lower success rates are not caused by taking fainter targets.
Also, even if we limit the $i'-J$ condition of W4\_002 ($i'-J>0.0$) to
that of W1\_002 ($i'-J>0.5$), the success rate of W4\_002 remains
constant ($11/119=9.2\%$). A more strict color selection of
  $g'-r'<0.5$ instead of $g'-r'<0.7$ for the photo-$z$ selections
  increases the overall success rate from $16.7\%$ to $19.1\%$, though
  the target number density drops to $\sim290$ per FoV.

We tested whether the success rate increases when $K$-band data
are included. All of the fields of W1\_004 and W4\_004 (NIR,
photo-$z$ selection) are covered also by the $K$ band, and about
60\% of CFHTLS Wide galaxies are detected both in the $J$ and $K$
band.  We then repeated the same target selection procedures given
by \S\ref{estimation} but with the $K$ band newly added.  Note that
a complete test of adding $K$ is impossible because the FMOS
observation is not available for galaxies that were not selected as
targets by the $u'g'r'i'z'J$ photo-$z$ selection, though some of
them would have been selected if we used also the $K$ band.
Therefore we simply examine the line detection rate in the
FMOS-observed galaxies that are also selected by the criteria
including $K$.  The success rate averaged over the W1 and W4 fields
then changes from 15.2\% to 15.6\%, i.e., almost no improvement. It
should be noted that these statistics are based on all the
FMOS-observed targets including those undetected in $J$ or $K$. If
we examine the galaxies detected in NIR bands, the success rate
becomes 16.7\% (optical+$J$ selection, detected in $J$) and 20.1\%
(optical+$JK$ selection, detected both in $J$ and $K$), and hence we
see a modest improvement.

\subsection{Spectroscopic versus photometric estimates
of redshift and H$\alpha$ flux}
\begin{figure*}[t]
 \begin{center}
 \includegraphics[width=8cm]{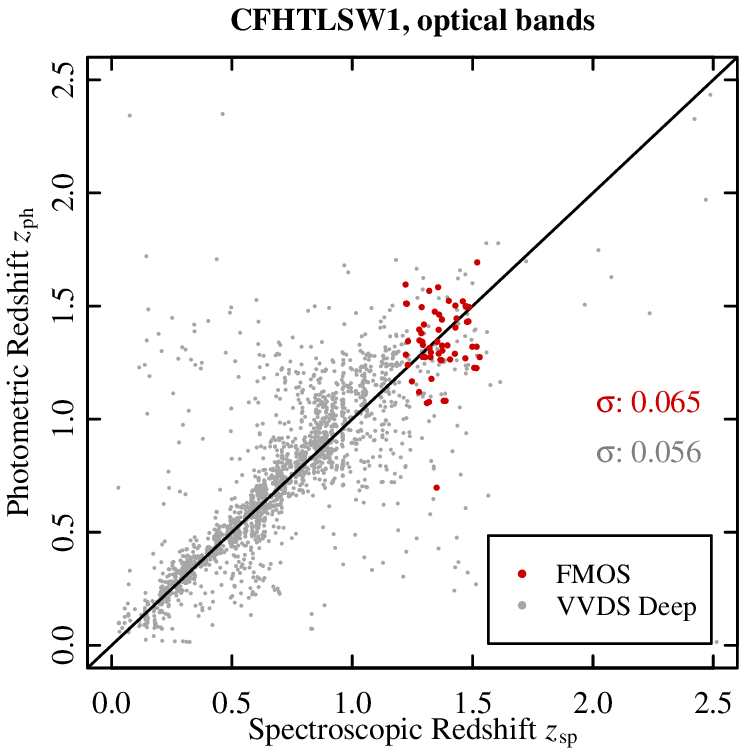}
 \includegraphics[width=8cm]{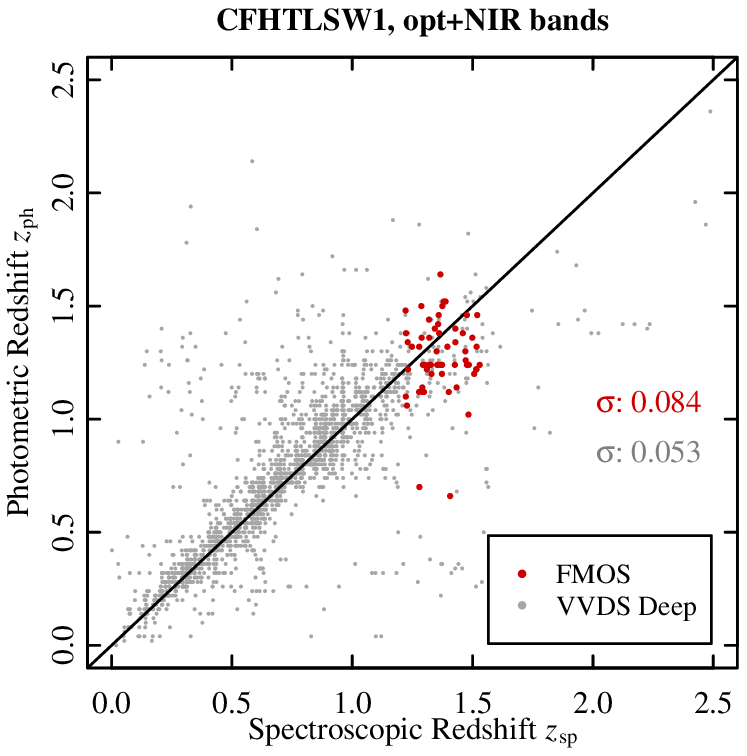}
 \includegraphics[width=8cm]{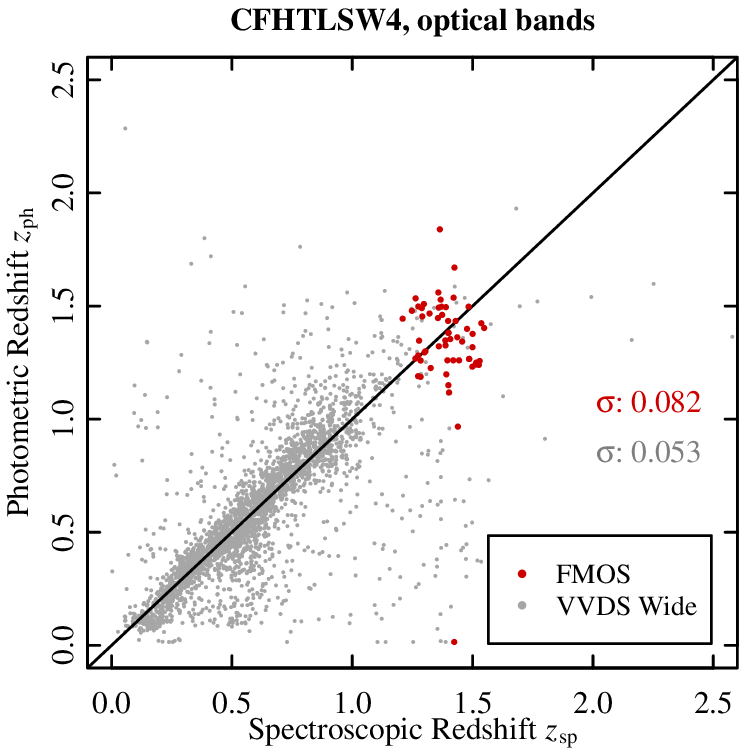}
 \includegraphics[width=8cm]{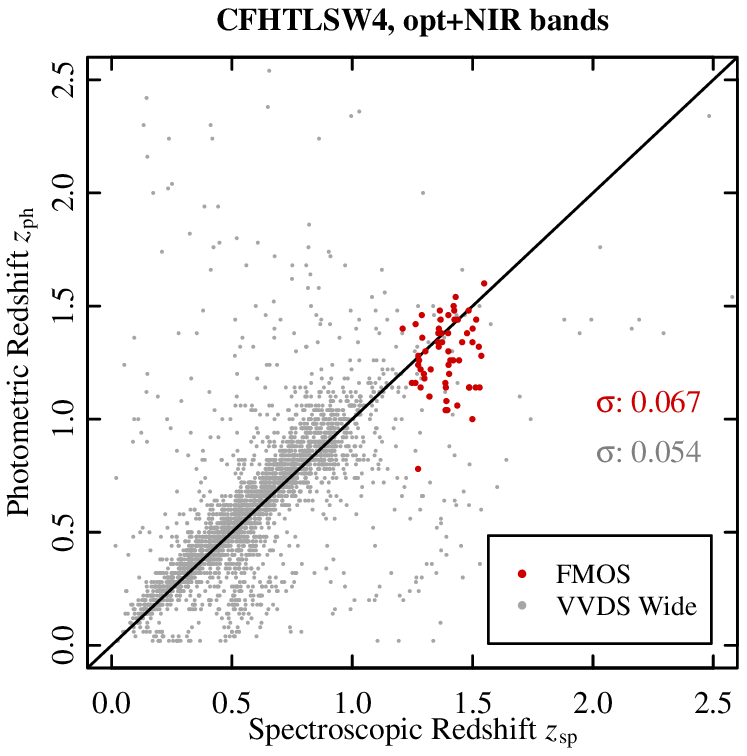}
 \caption{Comparison between spectroscopic and photometric redshifts
for galaxies with detected emission lines (red circles)
in the fields using
the photo-$z$ selection (W1\_003 and 004, and W4\_003 and 004). 
VVDS galaxies are also plotted for comparison.
$\sigma$ is the redshift accuracy calculated by $1.48\times
\mathrm{median}(\left|\Delta z\right|/(1+z_s))$.
}
 \label{comparez}
 \end{center}
\end{figure*}

\begin{figure}
 \begin{center}
 \includegraphics[width=8cm]{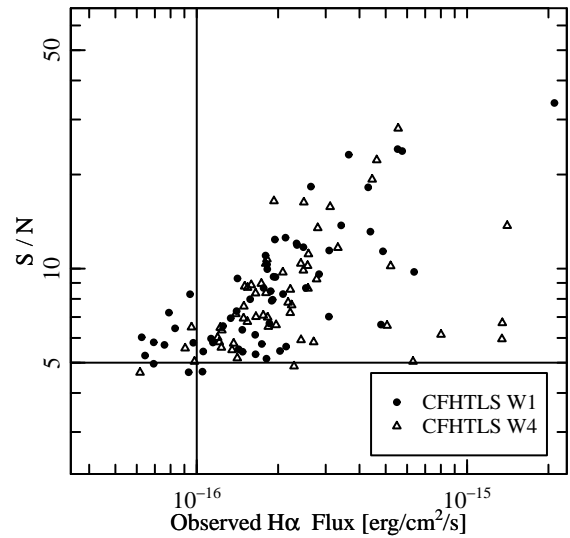}
 \caption{The signal-to-noise ratio of emission line detection 
plotted against the observed H$\alpha$ line fluxes.
  The observed fluxes are corrected for the effects of fiber aperture loss.
}
 \label{snobsha}
 \end{center}
\end{figure}

The observed spectroscopic redshifts are compared with photometric
redshifts in Figure \ref{comparez}. For comparison, the VVDS data used
for the photometric calibration of the redshift estimation are also
plotted in gray dots. 
$\sigma_{\Delta z/(1+z_s)}$ for galaxies observed by FMOS is larger ($\sim0.8$) than that for 
VVDS galaxies ($\sim0.6$). Note that $\sigma_{\Delta z/(1+z_s)}$ for FMOS galaxies is underestimated
due to the selection effect (i.e., we select target galaxies within the limited range of $z_{ph}$).
It should be noted that in our samples, adding
NIR data does not improve the redshift estimation nor decrease the
number of outliers, as discussed above.  A comparison between the line
$S/N$ and the observed H$\alpha$ flux is displayed in Figure
\ref{snobsha}. The H$\alpha$ flux corresponding to $S/N\sim5$ is $\sim
1.0\times10^{-16}\;{\rm [erg/cm^2/s]}$.

\begin{figure*}[t]
 \begin{center}
 \includegraphics[width=8cm]{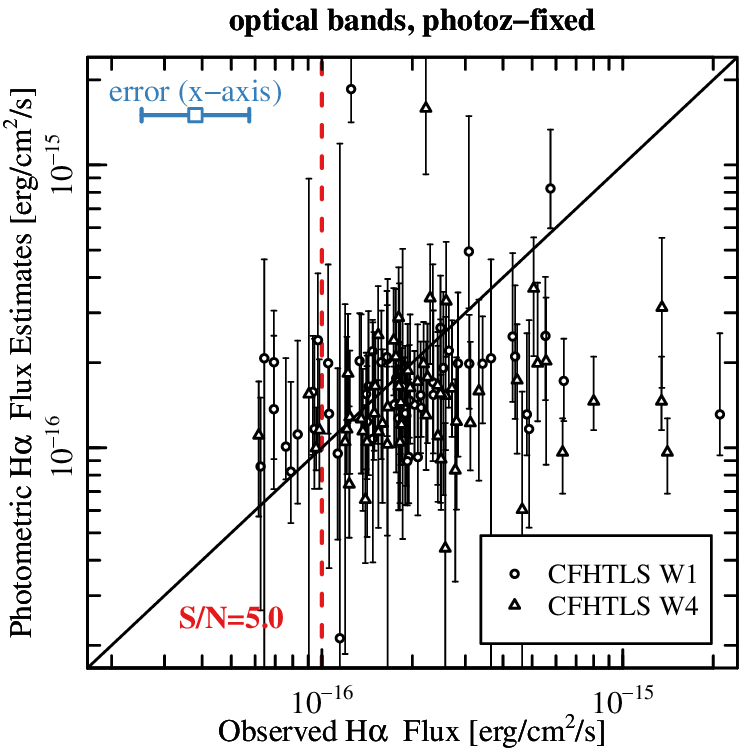}
 \includegraphics[width=8cm]{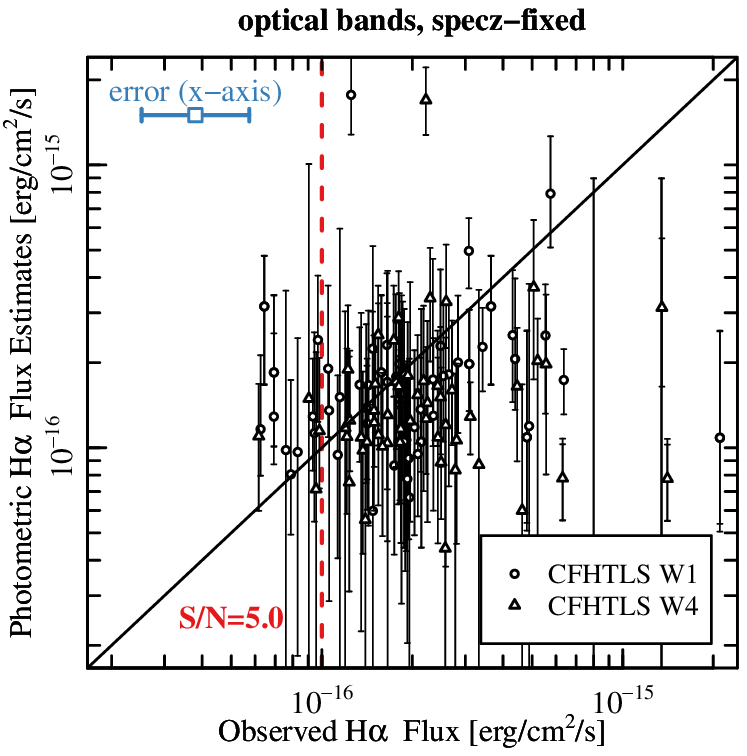}
 \includegraphics[width=8cm]{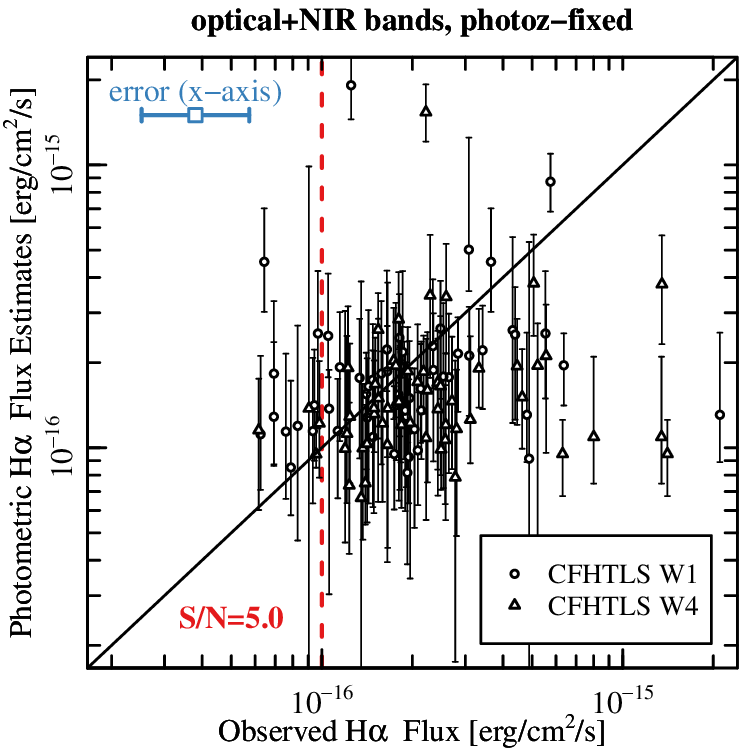}
 \includegraphics[width=8cm]{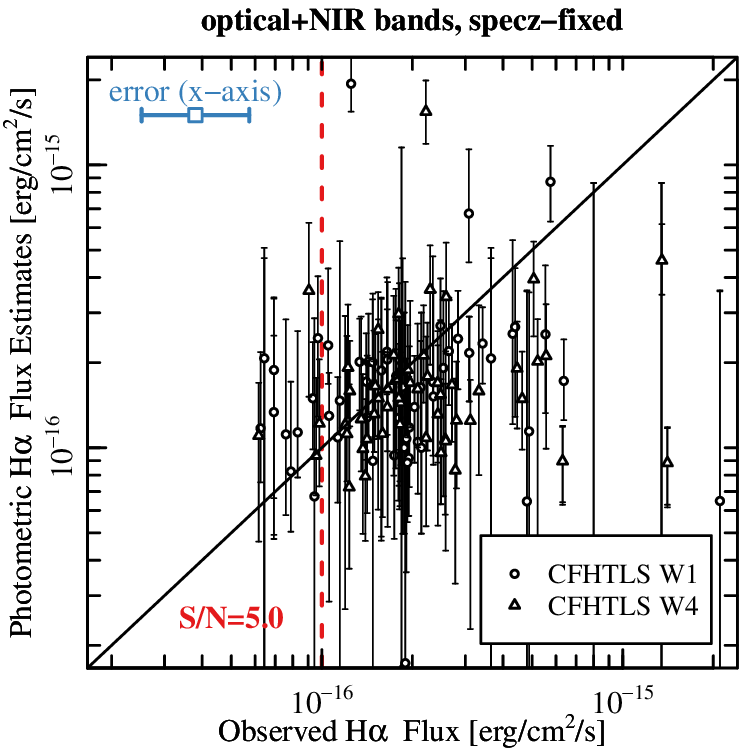}
 \caption{Comparison between the observed H$\alpha$ fluxes from FMOS
   and the fluxes estimated from photometric SED fitting.  Only
   galaxies in the FMOS fields with photo-$z$ selection (W1\_003, 004,
   W4\_003, 004) are shown.  The observed fluxes are corrected for the
   effect of fiber aperture loss (see \S\ref{reduction}).  The
   photometric H$\alpha$ flux estimates were made using optical bands
   \textit{(top panels)} or optical+NIR bands \textit{(bottom
     panels)}. In the left panels, redshifts are fixed at those
   estimated using SED fits to empirical SED templates, while in the
   right panels the redshifts were fixed at those determined
   spectroscopically. A typical error size along the $x$-axis
     by the fiber aperture loss (see \S\ref{detection}) is shown in
     the upper-left corner, and the flux level corresponding to $S/N =
     5$ is also shown. The $y$-axis errors are 1$\sigma$ statistical
     errors derived by the \textit{LePhare} code.}
 \label{compareha}
 \end{center}
\end{figure*}

The observed H$\alpha$ flux versus the estimated H$\alpha$ flux from
SED fittings is shown in Figure \ref{compareha}, which shows a large
scatter between the observed and estimated H$\alpha$ fluxes.  We
examined the effects of fiber loss variations for individual galaxy
sizes ($\sim\;50\%$) and by the positional error of fiber allocation
(typically less than 10\%, Yabe et al. 2012), and found that these are
not large enough to account for the scatter. Therefore this is
most likely to be caused by the uncertainties in the photometric
H$\alpha$ flux estimates.  The observed fluxes tends to be larger
than that of estimated fluxes, especially for those having large
observed fluxes. This is likely because of the selection effect
around the FMOS detection limit; most of galaxies are around the
selection threshold of the estimated flux, and some of them having
large observed fluxes have high probabilies of being detected.

In right panels of Figure \ref{compareha} we also plot the H$\alpha$
fluxes estimated by fixing redshifts in the SED fitting at the
measured spectroscopic values, rather than at the photometric
redshifts. There is still a large scatter, indicating that the theoretical
SED fitting has a large uncertainty, even if we use the correct
redshifts.  Note that we are observing in a relatively narrow redshift
range with FMOS for H$\alpha$ ($z=1.19$--$1.53$), and the selected
targets have photo-$z$ values in $z=1.18$--$1.54$. Therefore photo-$z$
and spec-$z$ cannot differ by large factors, and the lack of significant
improvement from using spec-$z$ should not be surprising.

\subsection{Comparison with the HiZELS Data}\label{hizels}
Here we examine our selection processes, with the aim of determining
which of the processes are mainly limiting the success rate of
emission line detection. We use data for emission line galaxies
detected by the NBH narrow-band filter of the High Redshift(Z)
Emission Line Survey (HiZELS: \cite{Geach}; \cite{Sobral2009}),
corresponding to H$\alpha$ at $z = 1.47$.  Although the redshift
interval corresponding to the narrow band filter is small
($z=1.466\pm0.016$), the redshift is within the FastSound
spectroscopic coverage, and all the H$\alpha$ emitter brighter than
the detection limit are included in the HiZELS sample.  Therefore we
can estimate how many H$\alpha$ emitters are missed by considering the
effects of the FastSound target selection criteria for the HiZELS
galaxies.  Note that this is entirely a calculation and we do not have
FMOS observations of HiZELS targets.

We use the HiZELS catalogue in a $0.63\;{\rm deg^2}$ region where the
UKIDSS UDS, SXDS, and CFHTLS W1 data are available, since the color
and photo-$z$ information from SXDS and the HiZELS NB921 narrow-band
imaging data in SXDS are crucial to classify HiZELS NBH emitters into
H$\alpha$ and others (\cite{Sobral}).  There are $192$ NBH emitters
having counterparts in the SXDS and CFHTLS W1 catalogues, while there
are $56{,}576$ galaxies in the same region in the CFHTLS wide
catalogue.  The NBH line flux limit is $7.0 \times
10^{-17}\;{\rm[erg/cm^2/s]}$ ($3\sigma$), which is close to the FMOS
line detection limit.

Among the $192$ objects, $114$ objects are identified to be H$\alpha$
emitters at $z \sim 1.47$ by the [OII]+H$\alpha$ double-line selection
using the HiZELS NB921 and HiZELS NBH data.  Since the NB921
observation is deep enough to detect galaxies with small emission line
ratios of [OII]/H$\alpha \sim 0.08$, more than $98\%$ of H$\alpha$
emitters can be selected by this criteria (\cite{Sobral2012}).
Considering the FMOS redshift coverage (OH mask regions removed)
  that is $8.6$ times wider than the HiZELS NBH width and the area of
FMOS FoV ($0.19\;{\rm deg^2}$), we expect $292$ H$\alpha$ emitters per
FMOS FoV, which is close to the number of FMOS fibers, indicating that
we need an almost perfect target selection method in order to achieve
$\sim\!100\%$ line detection efficiency.

NBH emitting objects that were not detected in NB921 should be
emission lines other than H$\alpha$, and these are useful to estimate
the contamination rate of non-H$\alpha$ emitters in the target
selection of FastSound.  Here we classify these into the following two
categories. One is the non-H$\alpha$ lines that are close to
H$\alpha$, such as [NII] $(6548, 6583)$ and [SII] $(6716, 6731)$.
Although H$\alpha$ emissions of these objects are outside the NBH
filter window, the redshift is almost the same as H$\alpha$ emitters,
and hence they can be detected as H$\alpha$ emitters by
FastSound. Therefore these should not be included in the contamination
rate estimate. Note that we do not have to count these as H$\alpha$
emitters either, since this population should be effectively included
in the above expected number of 375 H$\alpha$ emitters in the
FastSound redshift range.  The other category is the lines whose
redshifts are completely different, such as H$\beta (4861)$, [OIII]
$(4959, 5007)$, and [SIII] $(9069, 9532)$, which would cause a
redshift mis-identification and lead to a systematic error in $f\sigma_8$.

We classify the non-H$\alpha$ objects into these two categories accodring to
whether they satisfy photo-$z$ and color selection criteria for
H$\alpha$ identification of the NBH emitters adopted by the HiZELS
team (\cite{Sobral}).  Since the color/photo-$z$ criteria
does not discriminate between H$\alpha$ and the close lines, the
objects without NB921 emission but satisfying color/photo-$z$ criteria
are considered to be the former category.  There are $24$ and $54$
objects in the two categories, respectively, and we use only the
latter for the non-H$\alpha$ contamination rate estimate for
FastSound. Note that the actual contamination rate may be smaller than this estimate,
because we can identify H$\beta (4861)$+[OIII] $(4959,5007)$ double-emitters
and remove them in the actural FMOS spectra.

\subsubsection{Color-based Selection}\label{checkcolor}

\begin{table*}[t]
\begin{center}
  \caption{The statistics of HiZELS and CFHTLS galaxies after
    selection using optical colors described in \S\ref{checkcolor}.  The
    numbers in parentheses are scaled to the expected number in one
    FMOS field (0.19deg$^2$) and in the FastSound redshift coverage
    ($z=$1.19--1.53).  }
\begin{tabular}{lcccc}
\hline
\hline
Selection & HiZELS H$\alpha$ & HiZELS non-H$\alpha$ & CFHTLS & Success Rate\\
\hline
without selection & 114 (375) & 54 (178) & 56,576 (17,063) & 1.7\% \\
1a (color) & 21 (69) & 7 (23) & 4,742 (1,430) & 3.8\% \\
2a (1a + $g'$ mag) & 18 (59) & 2 (6.6) & 1,179 (356) & 13.1\% \\
3a (2a + $z'$ mag) & 17 (55) & 1 (3.3) & 1,100 (332) & 13.3\% \\
\hline
\hline
\end{tabular}
\label{SRcolor}
\end{center}
\end{table*} 

\begin{table*}
\begin{center}
  \caption{The same as Table \ref{SRcolor}, but for the case of
    optical photo-$z$ selection described in \S\ref{checkphotoz}.  }
\begin{tabular}{lcccc}
\hline
\hline
Selection & HiZELS H$\alpha$ & HiZELS non-H$\alpha$ & CFHTLS & Success Rate\\
\hline
without selection & 114 (375) & 54 (178) & 56,576 (17,063) & 1.7\% \\
1b ($z_{\rm ph}$) & 42 (138) & 11 (36) & 11,993 (3,560) & 3.0\% \\
2b (1b + H$\alpha$ flux) & 22 (72) & 4 (13) & 1,943 (586) & 9.7\% \\
3b (2b + $g'-r'$ color) & 21 (69) & 4 (13) & 1,599 (482) & 11.3\% \\
4b (3b + $z'$ magnitude) & 16 (53) & 4 (13) & 1,428 (431) & 9.6\% \\ 
\hline
\hline
\end{tabular}
\label{SRphotoz}
\end{center}
\end{table*}

\begin{figure*}
 \begin{center}
 \includegraphics[width=8cm]{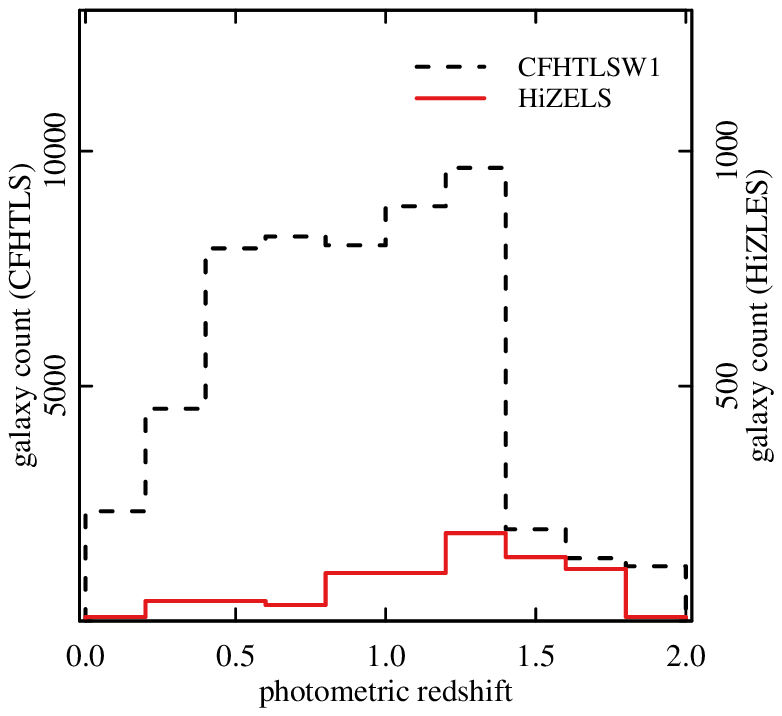}
 \includegraphics[width=8cm]{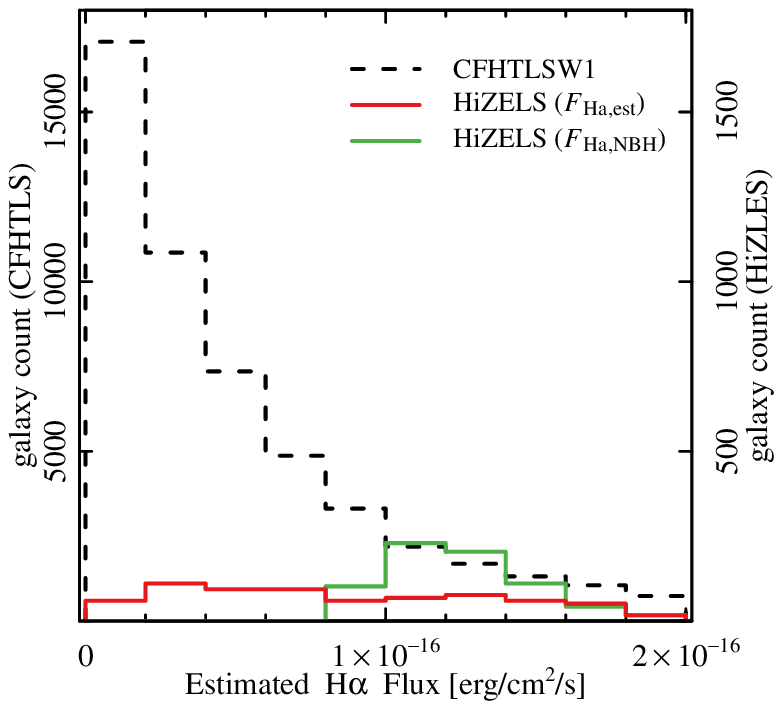}
 \caption{The distribution of photometric redshifts and photo-$z$
   based estimates of H$\alpha$ fluxes, for the HiZELS galaxies identified
   as H$\alpha$ emitters at $z = 1.47$ (H$\alpha$ fluxes brighter than
   $7.0 \times 10^{-17}\;{\rm[erg/cm^2/s]}$) and general CFHTLS
   galaxies. The number of HiZELS galaxies is multiplied by 8.6 to
   convert into the expected number in the FastSound redshift range.
   The actual H$\alpha$ fluxes measured by the HiZELS NBH band
   data are also shown by green blocks.}
 \label{hist}
 \end{center}
\end{figure*}

We first examine the color selection described in \S\ref{color} using
optical data, i.e., (1a) the $g'r'i'z'$ color conditions, (2a) the
$g'$ magnitude condition, and (3a) the $z'$ magnitude condition.
Here we adopt the $z'$ magnitude threshold of $z' < 23.0$, so that there are $\sim
400$ target galaxies in one FMOS field. After adopting the selection
criteria, $1{,}100$ target galaxies remain in the region of the HiZELS
data set, and there are $17$ HiZELS H$\alpha$ emitters and $1$
non-H$\alpha$ object. This means an H$\alpha$ detection rate of
$13.3\%$ if the redshift range is scaled to that of FMOS ($8.6$ times larger),
which is roughly consistent with the actual FMOS observation
reported in \S\ref{detectionrate}. In table \ref{SRcolor}, we also
show the change of galaxy numbers by adopting the each step of the
target selection of (1a)--(3a). The success rate (i.e., the number of
H$\alpha$ emitters in selected galaxies) increases by a factor of two
using the color condition and three using magnitude conditions, indicating that
conditions of color and magnitude are comparably important.

There is only one non-H$\alpha$ emitter in the final targets selected,
indicating that the contamination of non-H$\alpha$ in emission lines
of color selection is less than $10\%$, although the statistical uncertainty
is large.

\subsubsection{Photo$z$-based Selection}\label{checkphotoz}

Next we examine the optical photo-$z$-based selection, i.e., the
conditions on (1b) photometric redshift, (2b) estimated H$\alpha$
flux, (3b) $g'-r'$ color, and (4b) $z'$ magnitude.  For this region we
set $F_{\rm H\alpha, \ est}$ threshold to be $1.0 \times
10^{-16}\;{\rm [erg/cm^2/s]}$ to have an appropriate number of targets
($\sim\!400$) in one FMOS field. After these conditions have been adopted, $1{,}428$
CFHTLS galaxies and $16$ HiZELS H$\alpha$ emitters are selected,
yielding a success rate of $9.6\%$ scaled into the FastSound redshift
range, which is again roughly consistent with the FMOS
observations. The change of the success rate by the each step of the
selection (1b)--(4b) is also shown in Table \ref{SRphotoz}. This result
indicates that the most efficient selection process is obtained by adopting the
H$\alpha$ threshold, followed by the selection using photometric
redshifts.  The empirically introduced conditions of $(g'-r')$ color
and $z'$ magnitude are not very important at least for this sample.
 
The distribution of photometric redshifts and photo-$z$ based
estimates of H$\alpha$ fluxes for HiZELS H$\alpha$ emitters are
displayed in Figure \ref{hist}, in comparison with those for general
CFHTLS galaxies.  It is indeed seen that the fraction of HiZELS
H$\alpha$ emitters in general CFHTLS galaxies increases by choosing
galaxies at $z_{\rm ph} \sim 1.5$ and with strong estimated H$\alpha$
fluxes. However, considerable fraction of HiZELS H$\alpha$ emitters are
rejected by the adopted selection criteria, because of the
uncertainties in the estimated redshifts and
H$\alpha$ fluxes of HiZELS H$\alpha$ emitters which are at $z=1.47$
and brighter than $F_{\rm H\alpha}\sim0.7 \times10^{-16}\;{\rm [erg/cm^2/s]}$),
as seen in \S\ref{hist}. The accuracy of redshift
estimation against the correct value (i.e., $z = 1.47$) is
$\sigma_{\Delta z/(1+1.47)} \sim 0.13$, which is $\sim\!2$ times worse
than those reported for CFHTLS galaxies (Ilbert et al. 2006; Coupon
et al. 2009; see also \S\ref{estimation}).  This is because the
objects at $z\sim1.47$ are faint ($i'>22.0$) in the CFHTLS wide
catalogue, and because the accuracy is generally worse for blue
(H$\alpha$ emitting) galaxies than for red galaxies.

The number of non-H$\alpha$ emitters in the final selected targets is
relatively high: $4$ against $16$ H$\alpha$ emitters, implying a
contamination rate of $20\%$ that is considerably higher than that for
the color selection in the previous section.  However, by examining
the colors of the four non-H$\alpha$ objects, we found that three of
them are close to the border between H$\alpha$ and non-H$\alpha$ of
the HiZELS criteria in the color diagrams, and furthermore we can remove
H$\beta(4861)$+[OIII] $(4959, 5007)$ double-emitters using FMOS spectra,
hence the actual contamination may be reduced. A statistically larger sample is
required to investigate this issue more quantitatively.

\section{Conclusion}\label{conclusion}

In this work we examined several methods to efficiently select
H$\alpha$ emitting galaxies at $z \sim 1.2$--$1.5$ from existing
photometric data in a field of square-degree scales, using near-infrared
spectroscopic observations from Subaru/FMOS.  Although
quantitatively this work considers the particular case of the FastSound
project (a near-infrared galaxy redshift survey with FMOS), selecting
emission line galaxies efficiently at high redshifts is
important for future high-precision cosmological surveys that
are now widely being discussed.

The main photometric data set that we used is the CFHTLS wide survey,
where $u^*g'r'i'z'$ photometric data are available in four independent
fields, over a total of $155\;{\rm deg^2}$.  In order to test the gain
from adding near-infrared data, we also added $J$-band data from the
UKIDSS/DXS survey.  We tested two methods of target galaxy selection:
one is based on color-color diagrams, and the other is based on
photometric estimates of redshifts and H$\alpha$ fluxes from SED fitting
methods. Therefore we tested four types of target selections, by
combination of (1) color or photo-$z$ selections and (2) using optical
data only or optical plus NIR $J$ band.  Eight FMOS fields were
observed, corresponding to the four types of target selection in each
of the two fields of CFHTLS W1 and W4.

Emission lines are searched for by an automatic detection software
from the reduced spectra, and we calculated the success rates of
emission line detection ($S/N > 4.5$) in galaxies observed by
FMOS. The results are: $11.2\%$ (color, opt), $18.0\%$ (color,
opt+NIR), $17.0\%$ (photo-$z$, opt), and $15.7\%$ (photo-$z$, opt+NIR)
for CFHTLS W1, and $11.5\%$ (color, opt), $9.2\%$ (color, opt+NIR),
$17.7\%$ (photo-$z$, opt), and $14.6\%$ (photo-$z$, opt+NIR) for
CFHTLS W4 (summarized in Table \ref{rate}).  Though the sample size is
not  large enough for robust statistics, we find a tendency for higher success rates by
using photo-$z$ ($15$--$20\%$), relative to color selections
($\sim\!10\%$).  Adding NIR $J$ data does not appear to significantly
improve the success rate.

We also examined the target selection methods using the HiZELS NBH
narrow-band galaxy survey catalogue that corresponds to H$\alpha$ at
$z=1.47$.  We classified HiZELS sources into H$\alpha$ and
non-H$\alpha$ lines, and then applied our target selections to them,
in order to clarify what is the main factor limiting the success
rate, and how many non-H$\alpha$ lines are included as a
contamination in the FMOS-detected emission lines. We found that in the photo-$z$
selection, both the uncertainties of photo-$z$ and estimated H$\alpha$
flux are contributing to the relatively low success
rate.
%We also found that the contamination of non-H$\alpha$ lines in
%all the emission line galaxies detected by FMOS is $\sim\!10\%$,
%although further investigations with larger samples are neeeded.

\bigskip

We would like to thank David Sobral, Philip Best, Ian Smail, and
HiZELS team for providing us with the data obtained by the HiZELS
project.
This work is based on data collected at Subaru Telescope,
which is operated by the National Astronomical Observatory of Japan.
This work was supported by JSPS KAKENHI Grant Numbers 20040005,
22012005, and 23684007.

\end{document}